\documentclass[10pt]{iopart}
\usepackage{iopams}
\usepackage[pdftex]{graphicx}
\usepackage{epstopdf}
\usepackage{wrapfig}
\usepackage{bibentry}
\usepackage{multicol}
\usepackage{todonotes}
\usepackage{xcolor}
\usepackage{enumerate}

\newcommand{\NE}{$\left<n_e\right>$}
\definecolor{darkgreen}{rgb}{0.0,0.5,0} 
\definecolor{deepmagenta}{rgb}{0.8, 0.0, 0.8}
\definecolor{matlabred}{rgb}{0.851, 0.329, 0.102}
\definecolor{matlabblue}{rgb}{0, 0.451, 0.741}

\definecolor{72855}{rgb}{0.243, 0.149, 0.663}
\definecolor{72856}{rgb}{0.204, 0.478, 0.992}
\definecolor{72863}{rgb}{0.071, 0.745, 0.7255}
\definecolor{72861}{rgb}{0.784, 0.753,0.161}
\definecolor{72862}{rgb}{0.976,0.943,0.082}

\begin{document}
\title{Comparison of detachment in Ohmic plasmas with positive and negative triangularity}
\author{O. F\'evrier$^{1}$, C. K. Tsui$^2$, G.~Durr-Legoupil-Nicoud$^1$, C. Theiler$^1$, M. Carpita$^1$, S. Coda$^1$, C. Colandrea$^1$, B. P. Duval$^1$, S. Gorno$^1$, E. Huett$^1$, B. Linehan$^3$, A. Perek$^1$, L. Porte$^1$, H. Reimerdes$^1$, O. Sauter$^1$, E. Tonello$ ^1$, M. Zurita$^1$, T. Bolzonella$^4$, F. Sciortino$^5$, the TCV Team$^a$ and the EUROfusion Tokamak Exploitation Team$^b$}
\address{$^1$\'Ecole Polytechnique F\'ed\'erale de Lausanne (EPFL), Swiss Plasma Center (SPC), CH-1015 Lausanne, Switzerland.}
\address{$^2$Center for Energy Research (CER), University of California-San Diego (UCSD), La Jolla, California 92093, USA.}
\address{$^3$Plasma Science and Fusion Center MIT, Cambridge, Massachusetts 02139, USA.}
\address{$^4$Consorzio RFX, CorsoStati Uniti 4 – 35127 – Padova, Italy.}
\address{$^5$Max-Planck-Institut für Plasmaphysik, Boltzmannstraße 2, D-85748 Garching, Germany.}
\address{$^a$See the author list of H. Reimerdes et al 2022 \emph{Nucl. Fusion} \textbf{62} 042018.}
\address{$^b$See the author list of ``Progress on an exhaust solution for a reactor using EUROfusion multi-machines capabilities'' by E. Joffrin et al. to be published in Nuclear Fusion Special Issue: Overview and Summary Papers from the 29th Fusion Energy Conference (London, UK, 16-21 October 2023).}
\ead{olivier.fevrier@epfl.ch}

\begin{abstract}
In recent years, negative triangularity (NT) has emerged as a potential high-confinement L-mode reactor solution. In this work, detachment is investigated using core density ramps in lower single null Ohmic L-mode plasmas across a wide range of upper, lower, and average triangularity (the mean of upper and lower triangularity: $\delta$) in the TCV tokamak. It is universally found that detachment is more difficult to access for NT shaping. The outer divertor leg of discharges with $\delta\approx -0.3$ could not be cooled to below $5~\mathrm{eV}$ through core density ramps alone. The behavior of the upstream plasma and geometrical divertor effects (e.g. a reduced connection length with negative lower triangularity) do not fully explain the challenges in detaching NT plasmas. Langmuir probe measurements of the target heat flux widths ($\lambda_q$) were constant to within 30\% across an upper triangularity scan, while the spreading factor $S$ was lower by up to 50\% for NT, indicating a generally lower integral Scrape-Off Layer width, $\lambda_{int}$. The line-averaged core density was typically higher for NT discharges for a given fuelling rate, possibly linked to higher particle confinement in NT. Conversely, the divertor neutral pressure and integrated particle fluxes to the targets were typically lower for the same line-averaged density, indicating that NT configurations may be closer to the sheath-limited regime than their PT counterparts, which may explain why NT is more challenging to detach.
\end{abstract}

\maketitle

\ioptwocol
\section{Introduction} \label{sec:introduction}
To maximize fusion performance, it is generally considered that future reactors will operate in H-mode (\emph{High-confinement mode}), which generally features confinement times twice as high as for L-mode (\emph{Low-confinement mode}). However, H-mode operation is associated with challenges from the power exhaust perspective. To maintain H-mode, a high level of power crossing the separatrix is required. Such high power, combined with a narrower Scrape-Off Layer (SOL) width ($\lambda_q$) in H-mode compared to L-mode \cite{Eich_NF2013}, may result in divertor target power loads above the material limits. In particular, in the absence of mitigation, the target heat fluxes expected in ITER and DEMO are well above these limits \cite{Wischmeier_JNM2015,Reimerdes_NF2020_1}. Operation in a detached divertor regime is therefore foreseen for a fusion power plant \cite{Leonard_PPCF2018,Pitts_NME2019}. In this regime, most of the plasma exhaust heat is volumetrically dissipated by radiation and plasma-neutral interactions, while the total plasma pressure along the field lines in the SOL develops strong gradients, providing access to low target temperatures (below 5 eV) and lower particle fluxes at the targets \cite{Leonard_PPCF2018}. H-mode also generally comes with ELMs (Edge Localized Modes), that generate strong transient heat deposition at the divertor targets, possibly causing localized melting or other deleterious effects \cite{Pitts_NME2019}. While it is hoped that operation in a detached regime may (at least partially) buffer the ELM energy, or that ELM mitigation techniques (or operation in ELM-free scenarios) may alleviate this risk, operating in H-mode remains a strong challenge from the point of view of the power exhaust, and exploring different paths is thus important \cite{Viezzer_NME2023}.

One of these paths is Negative Triangularity (NT). Even in L-mode, NT discharges feature H-mode grade confinement without a steep edge pressure pedestal, and therefore without ELMs. This behavior was discovered in the TCV tokamak \cite{Pochelon_NF1999}, and further confirmed in experiments with ECRH heating \cite{Camenen_NF2007}, where NT discharges could sustain core pressure profiles similar to those of a matched PT discharge, but with half the auxiliary heating power. This behavior could be ascribed to a strong reduction of the turbulence levels in NT plasmas \cite{Fontana_NF2020}. Enhanced confinement in NT was also observed in the DIII-D tokamak \cite{Austin_PRL2019}, where NT discharges with $\beta_N=2.7$ and $H_{98,y2}=1.2$ were obtained, again in L-mode. In AUG, L-mode NT plasmas with high energy confinement were also observed in electron-heated plasmas \cite{Happel_NF2023}. A recent review of work on NT configurations can be found in \cite{Marinoni_RMPP2021}.

From a power exhaust perspective, operation in L-mode removes any need for ELM mitigation or ELM buffering. Furthermore, since H-mode no longer needs to be sustained, the power crossing the separatrix is no longer required to remain above the L-H threshold, thus allowing higher core and edge radiation scenarios. Synergistically, observations on the TCV tokamak show that negative triangularity discharges do not enter H-mode even with an input power significantly higher than the L-H threshold for positive triangularity discharges, therefore offering a wide L-mode parameter space for plasma operation in future reactors \cite{Coda_PPCF2022}. By operating in L-mode, the SOL width may be larger than in H-mode, which would be beneficial for power exhaust handling. Finally, the X-point is intrinsically at a higher major radius, that benefit the targets (increasing the target wetted area) without the expense of a big divertor volume or in-vessel poloidal field coils. From the reactor viewpoint, it is nevertheless likely that even for L-mode, operation with a divertor in a detached regime will be necessary. Therefore, it is crucial to study the access to this regime for negative triangularity configurations. 

The Tokamak à Configuration Variable (TCV) \cite{Reimerdes_NF2022} features a unique, highly flexible shaping capability thanks to 16 independently powered poloidal field coils, enabling the investigation of a large spectrum of core or divertor shapes \cite{Theiler_NF2017, Fevrier_NF2021}. Recently, the TCV team has undertaken a significant effort to explore further the parameter space of negative triangularity, in order to assess the possibility of using this configuration in the framework of a DEMO reactor \cite{Coda_PPCF2022}. In this paper, we explore negative triangularity from the point of view of power exhaust. In particular, we seek to characterize detachment in core density ramps. This paper is organized as follows. Section \ref{sec:setup} presents the experimental setup, that includes a description of the diagnostics used in this study. In section \ref{sec:detachment}, evidence is presented that NT plasmas are difficult to detach. Section \ref{sec:the_quest} presents experimental results to isolate the individual mechanisms that, combined, are thought to prevent detachment in NT. Finally, conclusions and future work are discussed in section \ref{sec:conclusion}.

\section{Experimental setup and diagnostic coverage}\label{sec:setup}
\begin{figure}[ht!]
\centering
\includegraphics[width=\linewidth]{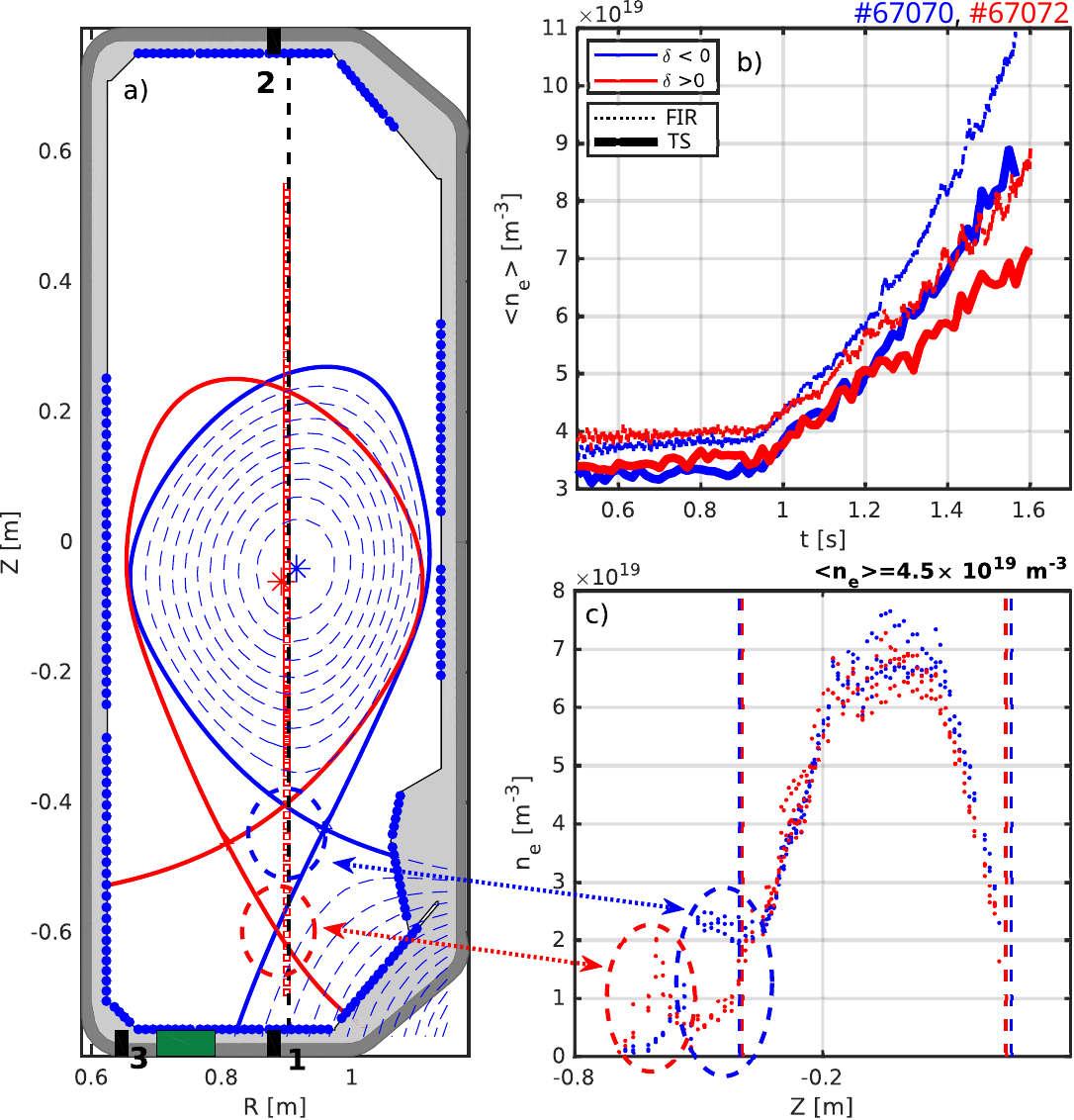}
\caption{\label{fig:TS_vs_FIR} (a) Separatrix shapes of NT (blue) and PT (red) configurations. The asterisks indicate the locations of the respective magnetic axes. The vertical dashed line corresponds to the sixth chord of the FIR, commonly used to define \NE\ (see \cite{Fevrier_NME2021} for instance). The red squares indicate the locations of the Thomson Scattering measurements. The blue dots correspond to the wall-embedded Langmuir probes that were available for all shots presented in this study. The numbered black rectangles at the bottom and top of the machine indicate the poloidal locations of the gas valves used for fuelling. The green rectangle shows the port onto which the divertor baratron pressure gauge is attached via an extension tube. (b) Typical time trace of the line-averaged densities determined from the sixth chord of the FIR (dashed) and from integrating the Thomson Scattering in the core region only (thick solid lines). (c) Electron density profiles for a given line-averaged density ($\left<n_e\right> = 4.5\times 10^{19}~\mathrm{m^{-3}}$, determined from Thomson Scattering in the core region), as a function of $Z$. The vertical lines indicate the location of the separatrix. The ellipses correspond to the ``regions of interest'' also highlighted in panel a).}
\end{figure}

This study was performed on TCV (major radius $R_0= 0.88~\mathrm{m}$, minor radius $a= 0.24~\mathrm{m}$, $B_0\approx 1.44~\mathrm{T}$). This work will focus on Ohmic, L-mode, diverted single null configurations such as the positive and negative triangularity examples shown in Figure \ref{fig:TS_vs_FIR}a.
In these experiments, unless specified otherwise, D$_2$ is injected at a fuelling rate controlled by a feedback loop based upon the line-integrated density measured by a vertical chord of a Far-Infrared interferometer (FIR). The radial positions of the gas valves are indicated by the numbered black rectangles. Wall-embedded Langmuir Probes (LPs) \cite{Fevrier_RSI2018,DeOliveira_RSI2019} are indicated by the blue dots, and vertical red squares indicate the locations of the Thomson Scattering (TS) measurements \cite{Arnichand_JoI2019}. The green rectangle shows the port on which the divertor baratron pressure gauge is attached to via an extension tube. In the following, the radial coordinate $\rho_\psi$ is the normalized poloidal magnetic flux, defined as $\rho_\psi=\sqrt{\left(\psi-\psi_0\right)/\left(\psi_1-\psi_0\right)}$ where $\psi$ is the poloidal magnetic flux, with $\psi_0$ and $\psi_1$ the flux at the magnetic axis and at the primary X-point, respectively.

Herein the triangularity, $\delta$, is taken as the mean between the top (or \emph{upper}) and bottom (or \emph{lower}) triangularity ($\delta_{top}$ and $\delta_{bot}$). Usually, the line-averaged core density \NE\ is calculated from the FIR by dividing the line-integrated measurement by the length of the portion of the beam within the LCFS, which assumes that the density outside of the LCFS is negligible. This is not always the case, in particular for NT, due to the presence of a high-density plasma region on the HFS of the X-point \cite{Coda_PPCF2022}. Therefore, we shall take the TS density, where \NE\ is defined as the integral of the TS-inferred density \emph{within} the LCFS, divided by the height of the plasma core at $R=0.9~\mathrm{m}$ (the radial location of the TS measurement points and one of the FIR chords). While this yields a \NE\ that is in qualitative agreement with the FIR measurement (Figure \ref{fig:TS_vs_FIR}b), a finite difference remains. In NT, a high-density region on the HFS of the X-point \cite{Coda_PPCF2022} clearly impacts the FIR measurements, resulting in an overestimated \NE, Figure \ref{fig:TS_vs_FIR}c. Further exploration of this high-density region is left for future work. We note, however, that this may impact the exhaust properties of NT configurations, for instance by promoting the detachment of the inner target or augmenting localized radiation around the X-point. 

\section{Comparing detachment in negative and positive delta discharges}\label{sec:detachment}
\begin{figure}[ht!]
\centering
\includegraphics[width=\linewidth]{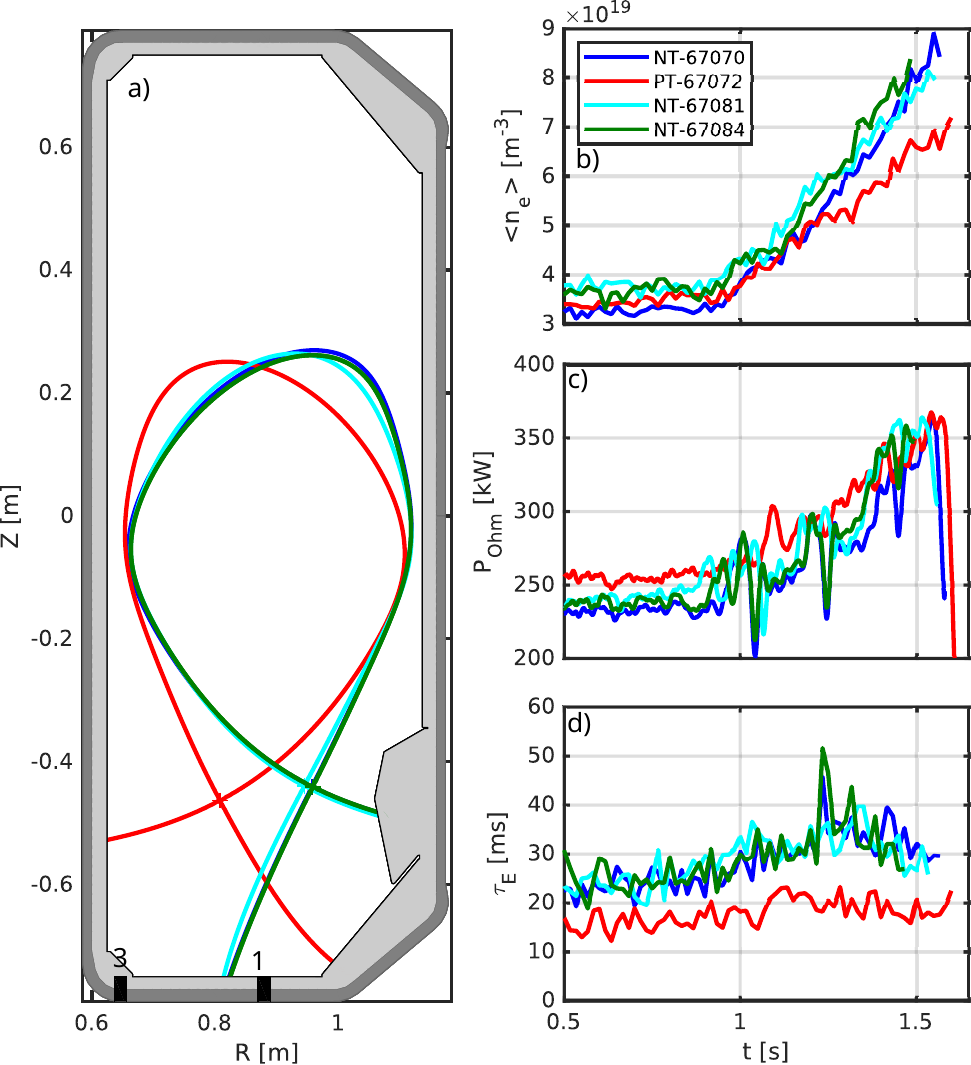}
\caption{\label{fig:basic_Scenario} (a) (blue and green) NT equilibrium ($\delta_{top}=-0.3$, $\delta_{bot}=-0.27$) (cyan) NT equilibrium ($\delta_{top}=-0.19$, $\delta_{bot}=-0.17$) (red) PT equilibrium ($\delta_{top}=0.27$, $\delta_{bot}=0.29$). The numbers indicate the gas valves, as defined in table \ref{tab:shotlist_densityscan} and Figure \ref{fig:TS_vs_FIR}a. (right panels) Time evolution of the (b) core line-averaged density from TS, (c) Ohmic power, and (d) energy confinement time, for the discharges in favorable field direction of table \ref{tab:shotlist_densityscan}.}
\end{figure}

In this section, we compare the NT and PT plasmas shown in Figure \ref{fig:basic_Scenario}a. These discharges are Lower Single-Null (LSN) with a plasma current $I_P = 225~\mathrm{kA}$. Experiments were performed both in ``favorable'' (for H-mode access) ion-$\nabla B$ direction (from the core towards the X-Point) and the opposite ``unfavorable'' $\nabla B$ direction. Table \ref{tab:shotlist_densityscan} presents a list of the discharges investigated in this section, together with some key parameters. Figure \ref{fig:basic_Scenario}b plots the evolution of \NE\ (from TS) for these discharges, with a density ramp commencing at $t=1$s. We observe that, for these scenarios, the NT cases disrupt at 25\% higher \NE\ than for PT. The Ohmic power, $P_{Ohm}$, is comparable for all discharges, albeit ~10\% higher for the PT case (Figure \ref{fig:basic_Scenario}b). All NT discharges exhibit a $\approx 50\%$ higher energy confinement time $\tau_E$ than the PT reference (Figure \ref{fig:basic_Scenario}c). 
\begin{table*}[h]
\begin{center}
\begin{tabular}{ c c c c c c c }
\hline
Discharge& $\delta_{top}$ & $\delta_{bot}$ & Field direction & fuelling location \\
\hline
\hline 
\textcolor{blue}{67070} & -0.3 & -0.27 & Favorable & V1 (Divertor, PFR) \\
\textcolor{cyan}{67081} & -0.17 & -0.19 & Favorable & V1 (Divertor, PFR) \\
\textcolor{darkgreen}{67084} & -0.3 & -0.28 & Favorable & V3 (Divertor, High Field Side CFR) \\
\textcolor{red}{67072} & 0.27 & 0.29 & Favorable & V1 (Divertor, PFR) \\
\hline
\textcolor{blue}{67465} & -0.3 & -0.27 & Unfavorable & V1 (Divertor, PFR)\\
\textcolor{red}{67467} & 0.27 & 0.29 & Unfavorable & V1 (Divertor, PFR)\\
\hline
\end{tabular}
\caption{{\label{tab:shotlist_densityscan} Summary of the plasma discharges used in section \ref{sec:detachment}. Fuelling is either done from the Common Flux Region (CFR) or from the HFS-Private Flux region (HFS-PFR).}}
\end{center}
\end{table*}

\subsection{Evolution of target parameters - Favorable field}
\begin{figure}[ht!]F
\centering
\includegraphics[width=\linewidth]{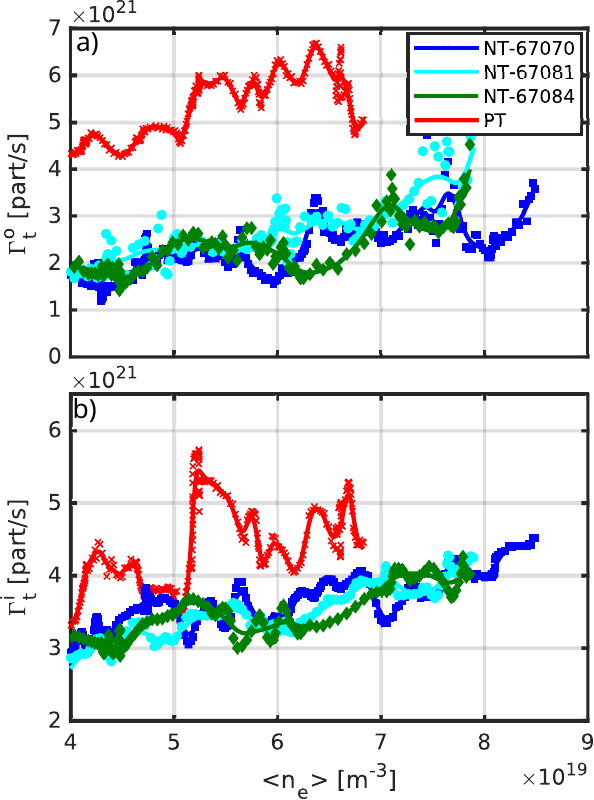}
\caption{\label{fig:Jsat_vs_neg_pos_neg} (a) Evolution of the integrated particle flux to the outer target $\Gamma_t^o$ and (b) to the inner target $\Gamma_t^i$ as a function of \NE, for the ``favorable field'' cases of table \ref{tab:shotlist_densityscan}.}
\end{figure}

\begin{figure}[ht!]
\centering
\includegraphics[width=\linewidth]{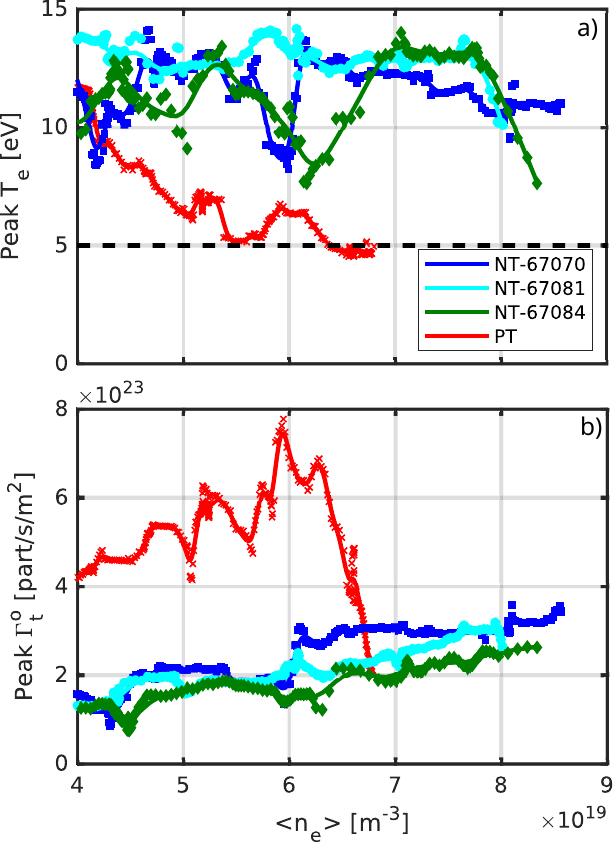}
\caption{\label{fig:Evolution_of_Te_Peak} (a) Outer target evolution of the peak electron temperature $T_e^{peak}$ (b) and of the peak parallel particle flux $J_{sat}^{peak}$ as a function of \NE, for the ``favorable field'' cases of table \ref{tab:shotlist_densityscan}. The dashed line in panel a) corresponds to the $T_e = 5~\mathrm{eV}$ threshold.}
\end{figure}

We first describe the outer target conditions for cases with a favorable field direction. The roll-over of the target ion flux is generally considered as an indicator for the onset of detachment \cite{Loarte_NF1998}. In the PT discharge, a roll-over of the integrated particle flux to the outer target, $\Gamma_t^o$, obtained from LPs, is seen at $\left< n_e \right> \approx 6.4\times 10^{19}~\mathrm{m^{-3}}$ (Figure \ref{fig:Jsat_vs_neg_pos_neg}a). No such roll-over is obtained in the NT discharges, where, instead, $\Gamma_t^o$ keeps increasing with \NE. This implies that, in the NT discharges, the outer target remains attached. Additionally, $\Gamma_t^o$ is lower in the NT cases compared to the PT case. A roll-over of the peak particle flux at the outer target, $J_{sat}^{peak}$, is only seen for the PT case, Figure \ref{fig:Evolution_of_Te_Peak}. The peak electron temperature at the outer target $T_e^{peak}$ decreases below the typical 5 eV threshold for a detached plasma only in the PT case, Figure \ref{fig:Evolution_of_Te_Peak}, confirming the integrated measurement in Figure \ref{fig:Jsat_vs_neg_pos_neg}. Similar conclusions can be drawn from the integrated particle flux to the inner target, $\Gamma_t^i$. In the PT discharge, signs of $\Gamma_t^i$ saturation are seen at $\left< n_e \right> \approx 5.5-6\times 10^{19}~\mathrm{m^{-3}}$ (Figure \ref{fig:Jsat_vs_neg_pos_neg}b), whereas $\Gamma_t^i$ keeps increasing through density ramp for NT. Despite different fuelling locations and/or slightly different triangularities, no strong changes are observed over the explored range of parameters among the NT cases, Table \ref{tab:shotlist_densityscan}.

\begin{figure*}[ht!]
\centering
\includegraphics[width=\linewidth]{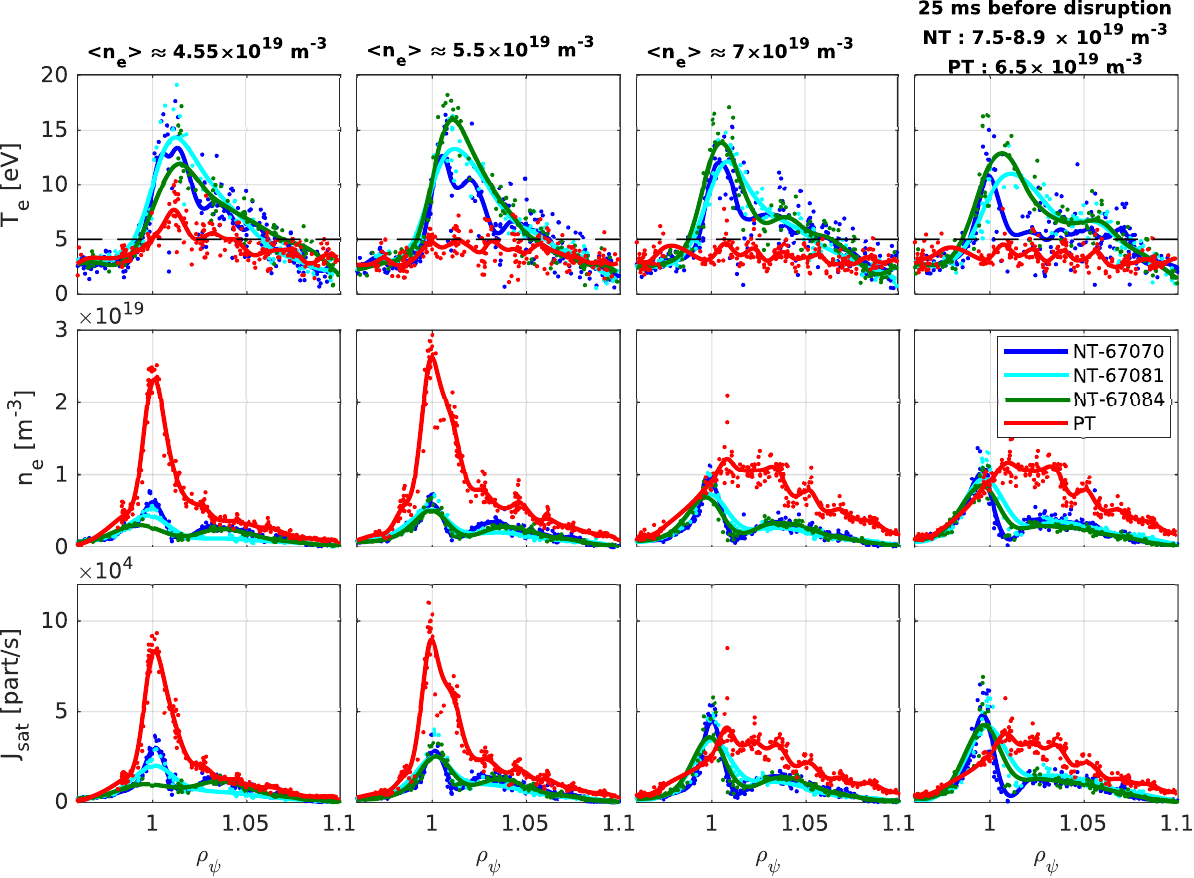}
\caption{\label{fig:Profiles_BeforeAfter} Outer target radial profiles of the electron temperature $T_e$ (top), electron density $n_e$ (middle) and parallel particle flux $J_{sat}$ (bottom) for different values of the line-averaged density \NE, for the ``favorable field" cases listed in table \ref{tab:shotlist_densityscan}.}
\end{figure*}

For more insight on the evolution of target profiles, Figure \ref{fig:Profiles_BeforeAfter} plots the ion saturation current density $J_{sat}$, electron density $n_e^t$, and temperature $T_e^t$ profiles from LP measurements at the outer target for different values of \NE. As \NE\ increases, both NT and PT cases exhibit a reduction of the peak temperature. However, in the NT cases, the reduction is modest with the peak temperature remaining above $5~\mathrm{eV}$ even at the highest density, whereas for PT, the temperature is strongly reduced across the entire profile. The three NT cases show very similar profiles. Together, these observations indicate that achieving detachment of the outer NT target is more difficult than for the corresponding PT discharge. 

\subsubsection{Influence of drifts' direction}
\begin{figure*}[ht!]
\centering
\includegraphics[width=\linewidth]{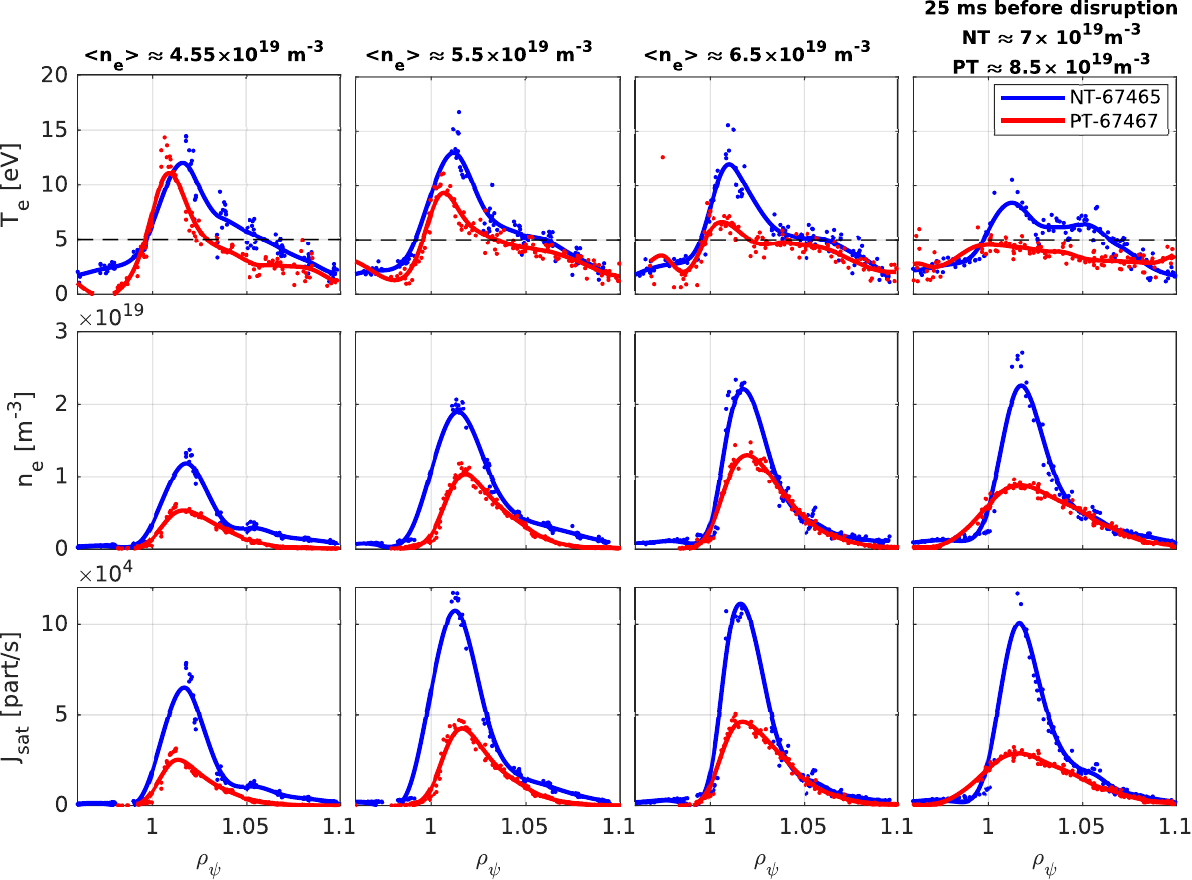}
\caption{\label{fig:drift_osp_profiles} Outer target radial profiles of the electron temperature $T_e$ (top), electron density $n_e$ (middle) and parallel particle flux $J_{sat}$ (bottom) for different values of the line-averaged density \NE, for a NT (blue) and a PT case (red), both with PFR fuelling and in ``unfavorable" field direction. On the top row, the thin black dashed line corresponds to the $T_e=5~\mathrm{eV}$ threshold.}
\end{figure*}

We next explore the impact of field direction, and hence the direction of drifts, on the previous observations. The drifts' direction is known to significantly affect the behavior of detachment, be it in the detachment threshold or in the dynamics of the transition to detachment \cite{Stangeby_NF1996,Potzel_NF2014,Jaervinen_PRL2018,Guo_NF2019,Fevrier_PPCF2020}. 
Figure \ref{fig:drift_osp_profiles} plots the ion saturation current density $J_{sat}$, target electron density $n_e^t$, and temperature $T_e^t$ profiles at the outer target for different values of \NE. The peak temperature, again, decreases with increasing \NE. However, as in the favorable field cases of the previous section, for NT cases, this reduction remains modest, with the peak temperature always remaining above $5~\mathrm{eV}$, whereas the temperature is strongly reduced across the entire profile for the PT case. Furthermore, for NT, both $J_{sat}$ and $n_e$ increase with increasing core density, whereas a roll-over of the peak of these quantities is observed for PT. Together, these observations indicate that the difficulty of achieving detachment in NT does not depend upon the field direction. 

\section{Isolating the mechanisms preventing detachment in negative triangularity}\label{sec:the_quest}
The findings from the preceding section show that density ramp detachment appears more difficult in NT than PT. We now investigate several mechanisms that could explain this observation. We first examine upstream quantities such as separatrix density and temperature, and the power to the SOL (section \ref{sec:upstream_quantities}). Second, we investigate any role of parallel connection length, and the ratio of inner to outer connection lengths (section \ref{sec:role_connlength}). We then examine the impact of divertor shape (section \ref{sec:role_divertorshape}). Additionally, we explore the role of the SOL width $\lambda_q$ (section \ref{sec:lambda_q}) and of the divertor neutral pressure (section \ref{sec:divertor neutral pressure}). 

\subsection{Evolution of upstream quantities}\label{sec:upstream_quantities}
\begin{figure}[ht!]
\centering
\includegraphics[width=\linewidth]{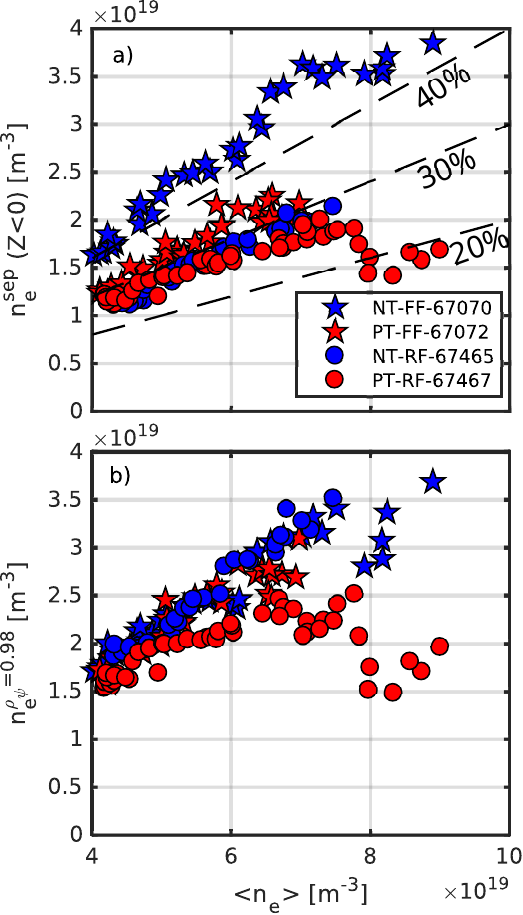}
\caption{\label{fig:edge_vs_nel_ne_TS} (a) Separatrix density, $n_e^{sep}$, and (b) Density at $\rho_\psi=0.98$, $n_e^{\rho_\psi=0.98}$ as a function of \NE, evaluated from the TS diagnostic for the cases of table \ref{tab:shotlist_upstreamparam}. Blue (resp. red) markers correspond to $\delta<0$ (resp. $\delta>0$) discharges. Stars indicate discharges in favorable field direction, whilst circles indicate the discharges in unfavorable field direction. The dashed lines in panel (a) indicate $n_e^{sep}=20\%\left<n_e\right>$, $n_e^{sep}=30\%\left<n_e\right>$, $n_e^{sep}=40\%\left<n_e\right>$. }
\end{figure}

We focus on the same discharges investigated in section \ref{sec:detachment}, table \ref{tab:shotlist_upstreamparam}. 
\begin{table*}[ht!]
\begin{center}
\begin{tabular}{ c c c c c c c }
\hline
Discharge& $\delta_{top}$ & $\delta_{bot}$ & Field direction & fuelling location \\
\hline
\hline
\textcolor{blue}{67070} & -0.3 & -0.27 & Favorable & V1 (Divertor, PFR) \\
\textcolor{red}{67072} & 0.27 & 0.29 & Favorable & V1 (Divertor, PFR) \\
\hline
\textcolor{blue}{67465} & -0.3 & -0.27 & Unfavorable & V1 (Divertor, PFR)\\
\textcolor{red}{67467} & 0.27 & 0.29 & Unfavorable & V1 (Divertor, PFR)\\
\hline
\end{tabular}
\caption{{\label{tab:shotlist_upstreamparam} Summary of the plasma discharges used in section \ref{sec:upstream_quantities}.}}
\end{center}
\end{table*}
One of the key parameters influencing the state of the divertor is the density at the separatrix, $n_e^{sep}$, taken here at the upstream location of the TS measurement, that is, at the intersection of the separatrix and the TS scattering measurement points, closest to the X-Point. Since this quantity is typically difficult to determine experimentally (due to uncertainties in TS measurements and in the equilibrium reconstruction), it is often substituted by \NE, as in the previous section. The question of using \NE\ as a proxy for $n_e^{sep}$ requires justification. Figure \ref{fig:edge_vs_nel_ne_TS}a) plots an evaluation of $n_e^{sep}$ from a linear fit through the TS profiles at $\rho=1$ for each of the NT and PT cases of table \ref{tab:shotlist_upstreamparam}, assuming that the equilibrium magnetic reconstruction is exact. For $\left<n_e\right> < 6.5\times10^{19}~\mathrm{m^{-3}}$, for NT and PT, the separatrix density is generally well represented by $n_e^{sep}\approx 0.25-0.30\times\left< n_e \right>$, which is consistent with previous observations on TCV \cite{Fevrier_NME2021, Wensing_PoP2021}. In the favorable field direction, NT case, however, $n_e^{sep}$ is generally higher with \NE\ ($n_e^{sep}/\left<n_e\right>\approx0.4$). This may result from the geometry of the TS diagnostic (Figure \ref{fig:TS_vs_FIR}), that will evaluate $n_e^{sep}$ at the HFS of the X-Point for NT cases, whereas it is at the LFS of X-Point for PT. For NT, the high-density region identified in Figure \ref{fig:TS_vs_FIR} may also increase the evaluated $n_e^{sep}$. Therefore, in Figure \ref{fig:edge_vs_nel_ne_TS}b), we present a second estimate of $n_e^{sep}$, based upon the measurement of the density at $\rho_\psi=0.98$, $n_e^{\rho_\psi=0.98}$, using the TS. For $\left<n_e\right> < 6.5\times10^{19}~\mathrm{m^{-3}}$, all cases have similar $n_e^{\rho_\psi=0.98}$. For $\left<n_e\right> > 6.5\times10^{19}~\mathrm{m^{-3}}$, the PT case with unfavorable field direction sees a decrease of $n_e^{\rho_\psi=0.98}$, as already observed for $n_e^{sep}$ , while the other cases still retain comparable $n_e^{\rho_\psi=0.98}$. For both estimations of the separatrix density, when $\left<n_e\right> > 6.5\times10^{19}~\mathrm{m^{-3}}$, $n_e^{sep}$ is overestimated by the formula $n_e^{sep}\approx 0.25-0.30\times\left< n_e \right>$ for PT discharges, that strengthens the observations of section \ref{sec:detachment}, since the PT actually detaches at even lower $n_e^{sep}$ than previously estimated. It is thus unlikely that differences in $n_e^{sep}$ can explain the observations in section \ref{sec:detachment}.

\begin{figure}[ht!]
\centering
\includegraphics[width=\linewidth]{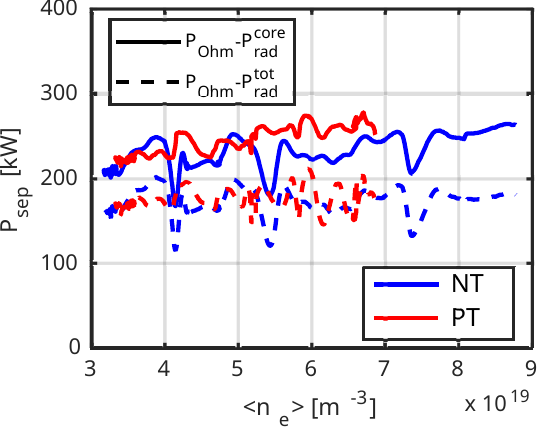}
\caption{\label{fig:Psep_POhm} Power crossing the separatrix, $P_{sep}$, for the NT case (blue) and the PT case (red) as a function of \NE. $P_{sep}$ is evaluated following both equations \ref{eq:psep_pradcore} and \ref{eq:psep_pradtot}.}
\end{figure}

Another key factor influencing the divertor state is the power crossing the separatrix, $P_{sep}$. The input power of these discharges, $P_{Ohm}$, is slightly higher for the PT discharge (Figure \ref{fig:basic_Scenario}c), ascribed to a lower core temperature (higher resistivity) for PT. The power crossing the separatrix is defined as 
\begin{equation}
P_{sep} = P_{Ohm} - P_{rad}^{core}
\label{eq:psep_pradcore}
\end{equation}
where $P_{rad}^{core}$ is the power radiated from the core, estimated from bolometry. Estimating the core radiated power is difficult due to the uncertainties in the bolometer tomographic inversion and the presence of a radiation region near the X-point that is difficult to classify simply either as ``core'' or ``SOL'' radiated power. Therefore, we also define a second expression for $P_{sep}$, $\tilde{P}_{sep}$, that assumes that all radiated power originates from the core.
\begin{equation}
\tilde{P}_{sep} = P_{Ohm} - P_{rad}^{tot}
\label{eq:psep_pradtot} 
\end{equation}
With these two estimations for $P_{sep}$, we find that the power crossing the separatrix is similar for both NT and PT (Figure \ref{fig:Psep_POhm}). It is, thus, unlikely that differences in $P_{sep}$ can explain the observations of section \ref{sec:detachment}.

\subsection{The role of connection length}\label{sec:role_connlength}
We now investigate the role of the divertor geometry. The PT configuration shown in Figure \ref{fig:basic_Scenario} has a relatively short inner leg (poloidally) and a longer outer leg, which is reversed for the NT configuration with a relatively short outer divertor leg. From the two-point model \cite{Stangeby_NF2018}, it is also expected that the parallel connection length will influence access to detachment, with longer parallel connection length promoting lower target temperature for given upstream conditions. Furthermore, it is well reported that the ratio of the parallel connection lengths of the two legs can influence the way the power is distributed between the two divertor targets \cite{Maurizio_NME2019}. Assuming the exhaust power reaches the divertor targets via parallel conduction (as in the two-point model \cite{Stangeby_NF2018}), the upstream temperature is considerably higher than the target temperature, $B\propto 1/R$ and no dissipation occurs along the flux tube, one can show that \cite{Maurizio_NME2019}
\begin{equation}
\frac{q_\|^o}{q_\|^i } = \frac{\mathcal{L}_\|^i}{\mathcal{L}_\|^o} \rightarrow \frac{q_\|^o}{q_\|^o + q_\|^i } = \frac{1}{1+\frac{\mathcal{L}_\|^o}{\mathcal{L}_\|^i} } 
\label{eq:share}
\end{equation}
where $q_\|^o$ (resp. $q_\|^i$) is the parallel heat flux to the outer (resp. inner) target and $\mathcal{L}_\|^o$ (resp. $\mathcal{L}_\|^i$) is the effective parallel connection length to the outer (resp. inner) target, defined as 
\begin{equation}
\mathcal{L}_\| = \int_0^{L_{\|}} \frac{R_u}{R\left(s_{\|}\right)}\mathrm{d}s_\|
\end{equation}
where $L_{\|}$ is the parallel connection length of the flux tube from an upstream location (taken here at the outboard midplane), from where the power enters the flux tube, to the target. $R\left(s_{\|}\right)$ is the major radial position along the flux tube, with $R_u$ the major radius at the upstream location. Such a change in power sharing with connection length was shown experimentally \cite{Maurizio_NME2019}. Typical NT configurations in TCV tend to exhibit lower parallel connection length to the outer target, $L_\parallel^{outer}$, while also decreasing the ratio ${\mathcal{L}_\|^o} / \mathcal{L}_\|^i$. That, according to equation \ref{eq:share}, results in an increase in ${q_\|^o}/\left\{q_\|^o+q_\|^i\right\}$, and, thus, an increase in the power to the outer target, together with a more difficult access to detachment. 

In practice, both $L_{\|}$ and $\mathcal{L}_\|$ diverge close to the separatrix. Therefore, in the following, we will define an averaged effective parallel connection length ratio, $\mathcal{R}$, by
\begin{equation}
\mathcal{R} = \frac{1}{4.95~\mathrm{[mm]}}\int_{0.5~\mathrm{[mm]}}^{5~\mathrm{[mm]}} \frac{\mathcal{L}_\|^o}{\mathcal{L}_\|^i} ~\mathrm{dR_{u}}
\end{equation}
where $dR_{u}$ corresponds to the distance to the separatrix, mapped upstream. $\mathcal{R}$ is thus an average of the ratio ${\mathcal{L}_\|^o}/{\mathcal{L}_\|^i}$ across the first $\approx 5$ mm of the SOL that avoids the divergence near the separatrix. We note that, by changing the geometry, both $\mathcal{R}$ and $L_{\|}$ are affected. These two quantities are indeed well correlated and difficult to disantangle in these experiments.

\subsubsection{Investigating the role of $\mathcal{R}$ and $L_{\|}$}\label{subsec:roleofR}
\begin{figure}[ht!]
\centering
\includegraphics[width=\linewidth]{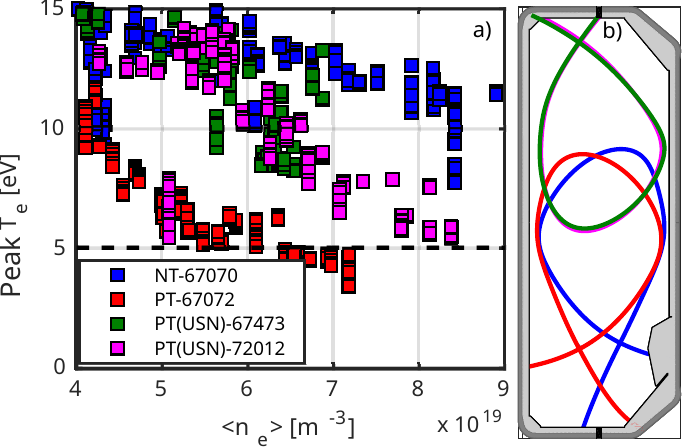}
\caption{\label{fig:PeakTS_vs_N} (a) Peak outer target electron temperature as a function of \NE\ for the cases reported in Table \ref{tab:shotlist_USN}. (b) Separatrix shape for a NT LSN case (blue), the PT USN cases (magenta, green), the PT LSN case (red) listed in table \ref{tab:shotlist_USN}. The black rectangles at the top and bottom indicates the location of the fuelling valves.}
\end{figure}

To obtain close comparative values of $\mathcal{R}$ and $L_{\|}$ between PT and NT discharges ($\mathcal{R}\approx0.27-0.37$, table \ref{tab:shotlist_USN}), we perform Upper Single-Null (USN) PT discharges, characterized by a poloidal length of the outer leg similar to the $\approx 11~\mathrm{cm}$ value of the NT case, Figure \ref{fig:PeakTS_vs_N}b. $\mathcal{R}$ and $L_{\|}$ are lower, Table \ref{tab:shotlist_USN}, sitting half-way between the values obtained in the NT and PT configurations of section \ref{sec:detachment}. In the USN discharges, the magnetic field direction has been reversed compared to the LSN counterparts so the ion-$\nabla B$ drift direction with respect to the divertor remains unchanged.

\begin{table*}[h]
\begin{center}
\begin{tabular}{ c c c c c c c c c c}
\hline
Discharge& $\delta_{top}$ & $\delta_{bot}$ & $I_P$ & $B\times\nabla B$ direction & Fuelling & Configuration & $\mathcal{R}$ & $L_\parallel^{outer}$ & $L_\parallel^{inner}$ \\
\hline
\textcolor{blue}{67070} & -0.3 & -0.28 & $225~\mathrm{kA}$ & Towards X-point & V1 (bottom) & LSN & $0.27$ & $6.8~\mathrm{m}$ & $22.1~\mathrm{m}$ \\
\textcolor{red}{67072} & 0.28 & 0.29 & $225~\mathrm{kA}$ & Towards X-point & V1 (bottom) & LSN & $0.54$ & $13.84~\mathrm{m}$ & $19.5~\mathrm{m}$ \\
\textcolor{darkgreen}{67473} & 0.31 & 0.29 & $225~\mathrm{kA}$ & Towards X-point & V1 (bottom) & USN & $0.37$ & $9.68~\mathrm{m}$ & $19.8~\mathrm{m}$ \\
\textcolor{deepmagenta}{72012} & 0.37 & 0.29 & $225~\mathrm{kA}$ & Towards X-point & V2 (top) & USN & $0.37$ & $9.44~\mathrm{m}$ & $19.2~\mathrm{m}$ \\
\hline
\end{tabular}
\caption{{\label{tab:shotlist_USN} Summary of the plasma discharges and key parameters used in section \ref{subsec:roleofR}. }}
\end{center}
\end{table*}
With intermediate values of $\mathcal{R}$ and $L_{\|}$, Figure \ref{fig:PeakTS_vs_N}, the outer target peak temperature of the USN cases lies in between the two LSN cases investigated in section \ref{sec:detachment}. This makes $\mathcal{R}$ and $L_{\|}$ good candidates to explain the difficulty in reaching detachment for the NT configurations of section \ref{sec:detachment}.

\subsubsection{Scanning the bottom triangularity}\label{subsec:scanbottomdelta}
\begin{figure}[ht!]
\centering
\includegraphics[width=\linewidth]{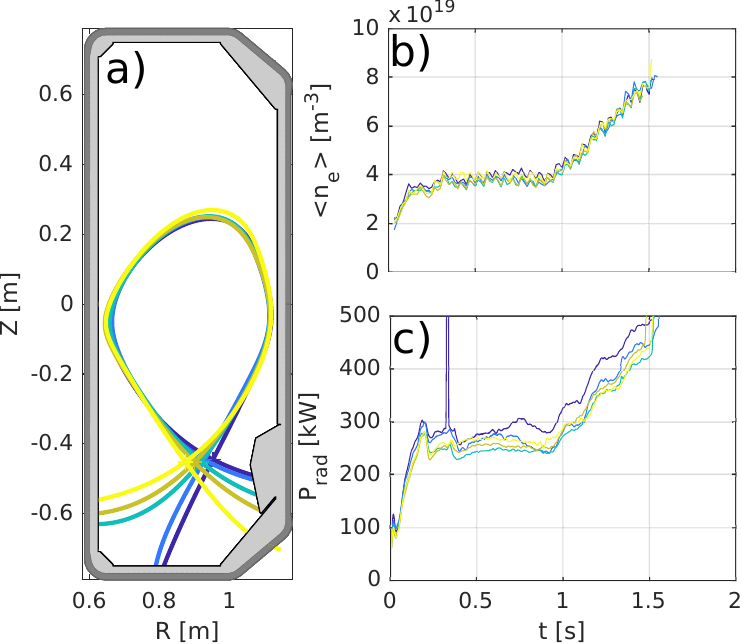}
\caption{\label{fig:bot_delta_scan_densities} (a) Shapes of the discharges listed in Table \ref{tab:summary_bot_delta_scan}. (b) Core line-averaged $\left< n_e \right>$ as a function of time for these discharges. (c) Radiated power $P_{rad}$ for these discharges}
\end{figure}

The effect of $\mathcal{R}$ and $L_{\|}$ is further probed in a scan of the bottom triangularity (varied from -0.27 to 0.05) whilst keeping the top triangularity as constant as possible ($\delta_{top}\approx -0.27$), Figure \ref{fig:bot_delta_scan_densities}a. The outer target poloidal flux expansion was reasonably constant over this scan, between $2-2.5$. Table \ref{tab:summary_bot_delta_scan} summarizes the discharges used, that were all in the ``favorable'' $\nabla B$ direction. Across these discharges, the radiated power was comparable (Figure \ref{fig:bot_delta_scan_densities}c). 

\begin{table}[h]
\center
\begin{tabular}{ l c c c c c c}
\hline
Discharge& $\delta_{top}$ & $\delta_{bot}$ & $\delta$ & $L_{\|}^o$ [m] & $\mathcal{R}$ \\
\hline
\textcolor{72855}{72855} & -0.27 & -0.28 & -0.27 & 6.4 & 0.23\\
\textcolor{72856}{72856} & -0.26 & -0.17 & -0.22 & 7.01 & 0.25\\
\textcolor{72863}{72863} & -0.25 & -0.14 & -0.19 & 8.18 & 0.31\\
\textcolor{72861}{72861} & -0.24 & -0.08 & -0.16 & 9.26 & 0.36\\
{\textcolor{72862}{72862}} & -0.26 & 0.05 & -0.11 & 10.62 & 0.44\\
\hline
\end{tabular}
\caption{{\label{tab:summary_bot_delta_scan} Summary of the main plasma discharges used in section \ref{subsec:scanbottomdelta}. $L_{\|}^o$ refers to the parallel connection length from the outer midplane to the outer target.}}
\end{table}

\begin{figure}[ht!]
\centering
\includegraphics[width=\linewidth]{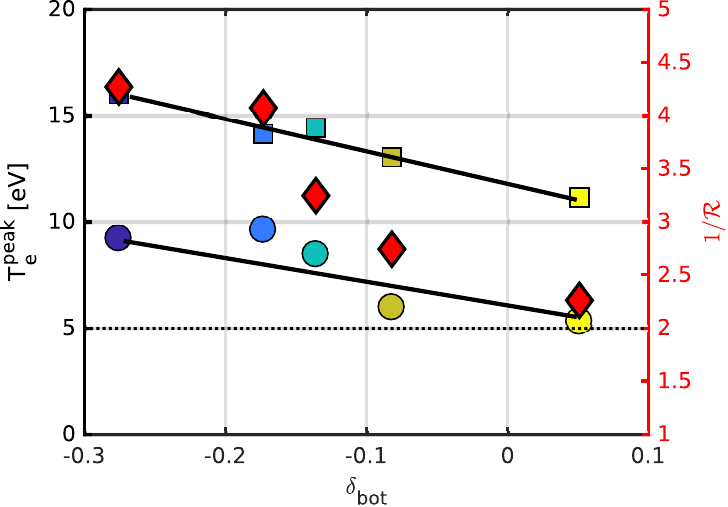}
\caption{\label{fig:bot_delta_scan_tepeak} (left axis) Peak electron temperature $T_e^{peak}$ (eV) at the outer target as a function of the bottom triangularity $\delta_{bot}$ for $\left< n_e \right> = 4\times10^{19}~\mathrm{m^{-3}}$ (squares) and $\left< n_e \right> = 7\times10^{19}~\mathrm{m^{-3}}$ (circles). The dashed line indicates $T_e^{peak} \approx5\mathrm{eV}$, while the two solid lines are guides for the eye. The same colormap is used in this Figure and in Figure \ref{fig:bot_delta_scan_densities} to identify the discharges, whose list can be found in Table \ref{tab:summary_bot_delta_scan}. (right axis) $1/\mathcal{R}$ vs $\delta_{bot}$ (diamonds). }
\end{figure}

At both low- ($\left< n_e \right> = 4\times10^{19}~\mathrm{m^{-3}}$) and high-density ($\left< n_e \right> = 7\times10^{19}~\mathrm{m^{-3}}$), the peak electron temperature $T_e^{peak}$ (eV) at the outer target decreased with increasing $\delta_{bot}$ (Figure \ref{fig:bot_delta_scan_tepeak}). This is compatible with the previous sections' findings (e.g. section \ref{sec:detachment}). One should, however, note that varying $\delta_{bot}$ changes both $L_{\|}$ and $\mathcal{R}$, as indicated in Table \ref{tab:summary_bot_delta_scan} and, for $\mathcal{R}$, in Figure \ref{fig:bot_delta_scan_tepeak}.

This section shows how geometrical effects, related to connection lengths (absolute and effective), can explain, at least partially, the difficulty in reaching detached conditions in the NT plasmas of section \ref{sec:detachment}. The next section will, however, demonstrate that even when the divertor shape is matched, the difficulty in detaching NT plasma partly persists, suggesting additional mechanisms are at play.

\subsection{The role of divertor shape}\label{sec:role_divertorshape}\label{sec:subsub_matchingdivertorshape}
\begin{figure}[ht!]
\centering
\includegraphics[width=\linewidth]{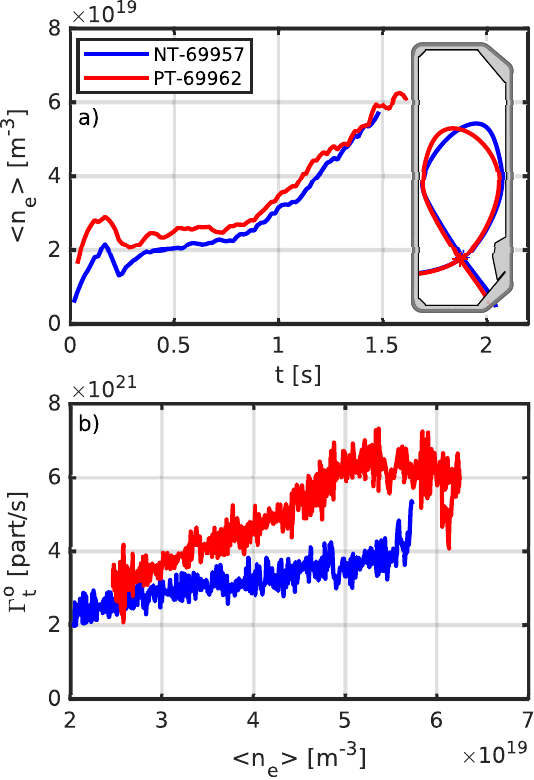}
\caption{\label{fig:Hybrid_summary} \label{fig:LP_OSP_integrated} (a) \NE\ determined from the TS in the core region only for the discharges listed in Table \ref{tab:hybrid_shapes}. The inset plots the NT and PT shapes. (b) Evolution of the integrated particle flux to the outer target $\Gamma_t^o$ as a function of \NE, for the PT case (red), and the NT case (blue).}
\end{figure}

\begin{figure*}[ht!]
\centering
\includegraphics[width=\linewidth]{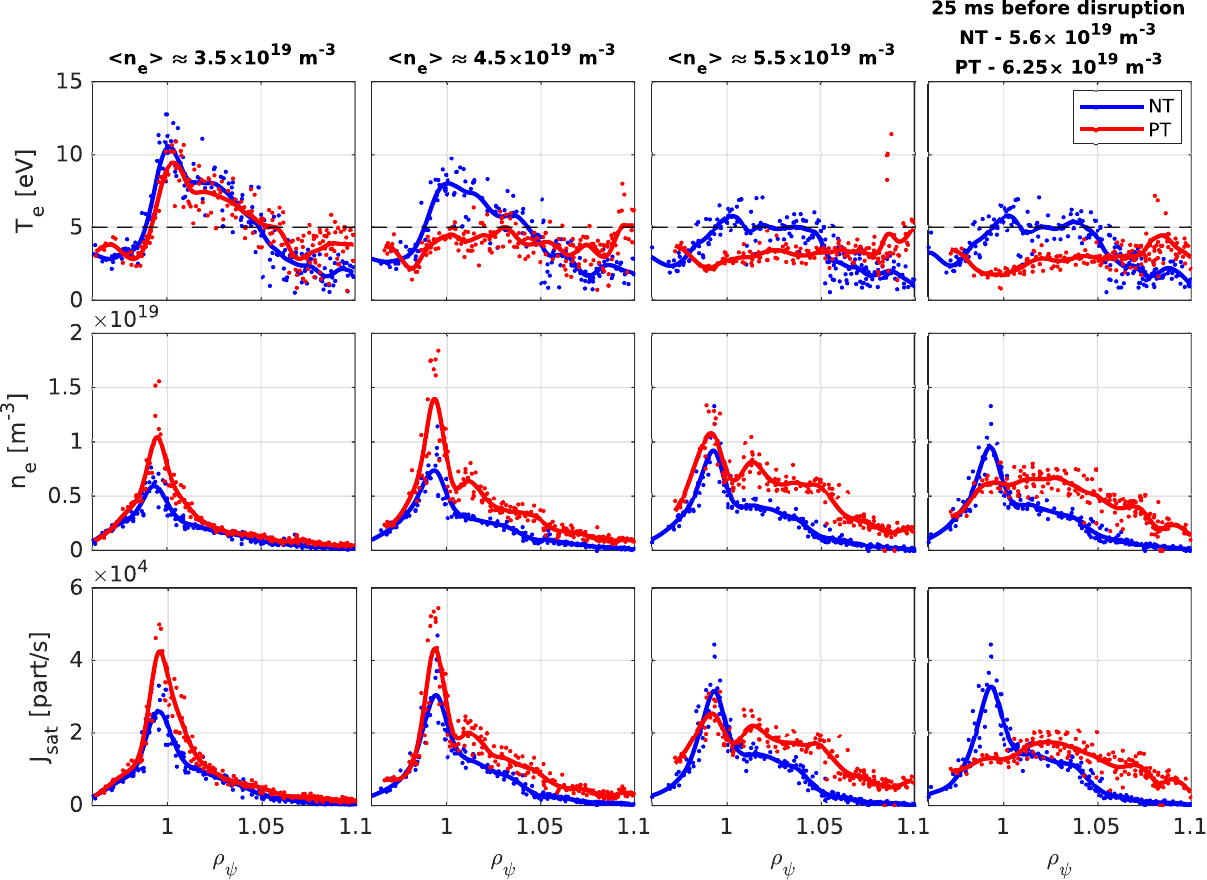}
\caption{\label{fig:Hybrid_LP_profiles} Radial $T_e$ (top row), $n_e$ (middle row), $J_{sat}$ (bottom row) profiles at the OSP for different values of \NE\ ($3\times 10^{19}~\mathrm{m^{-3}}$, $4.5\times 10^{19}~\mathrm{m^{-3}}$, $6\times 10^{19}~\mathrm{m^{-3}}$, 50 ms before the disruption), for the discharges in table \ref{tab:hybrid_shapes}.}
\end{figure*}

In previous sections, NT and PT configurations were compared with similar scenarios but with different divertor shapes. Here, we perform a comparison of NT and PT with matching divertor shapes. The scenarios investigated are \NE-ramps in LSN configurations, Figure \ref{fig:Hybrid_summary}a, with the magnetic field is in the ``favorable" direction. The lower triangularity is close to 0 ($\delta_{bot}\approx -0.02$) and $\delta_{top}$ is varied from negative ($\delta_{top}=-0.3$) to positive ($\delta_{top}=0.2$), Figure \ref{fig:Hybrid_summary}a. Table \ref{tab:hybrid_shapes} provides a summary of the relevant parameters. The parallel connection lengths to the outer target are similar ($L_{\|} = 14.7~\mathrm{m}$ for NT and $L_{\|} = 15.7~\mathrm{m}$ for PT), as are the $\mathcal{R}$ parameters (0.51 for NT and 0.54 for PT). The target flux expansions $f_x$ are also well matched ($f_x\approx 6.4$ at the outer target for both PT and NT). 
\begin{table}[h]
\center
\begin{tabular}{ l c c c c }
\hline
Discharge& $\delta_{top}$ & $\delta_{bot}$ & $\mathcal{R}$ & $L_{\|}^o$ [m]\\
\hline
\textcolor{blue}{69957} & -0.3 & -0.02 & 0.51 &14.7 \\
\textcolor{red}{69962} & 0.2 & -0.02 & 0.54 & 15.7\\
\hline
\end{tabular}
\caption{{\label{tab:hybrid_shapes} Summary of the main plasma discharges used in section \ref{sec:role_divertorshape}.}}
\end{table}
A clear roll-over in the integrated target ion flux is observed at the outer target ($\Gamma_t^o$) of the PT, at $\left<n_e\right> \approx 5.25\times 10^{19}~\mathrm{m^{-3}}$ (Figure \ref{fig:LP_OSP_integrated}b), indicative of detachment, whereas no roll-over was observed for NT. Figure \ref{fig:Hybrid_LP_profiles} plots the ion saturation current density $J_{sat}$, the target electron density $n_e$, and the temperature $T_e$ profiles from the LPs at the OSP for different values of \NE. All cases show a clear reduction in the peak $T_e$ as \NE\ is increased. However, while the PT target temperature drops to below 5 eV, indicative of detachment, it remains above 5 eV in the NT case, indicating that the plasma remained attached during the entire density ramp. The density profiles show an increase followed by a decrease of the peak density for PT, that is expected for a detaching plasma. For NT, however, the peak density initially increases, and then saturates at $n_e \approx 10^{19}~\mathrm{m^{-3}}$. This is supportive of a higher degree of detachment of the outer target in the PT case, compared to the NT case. 

\begin{figure*}[ht!]
\centering
\includegraphics[width=\linewidth]{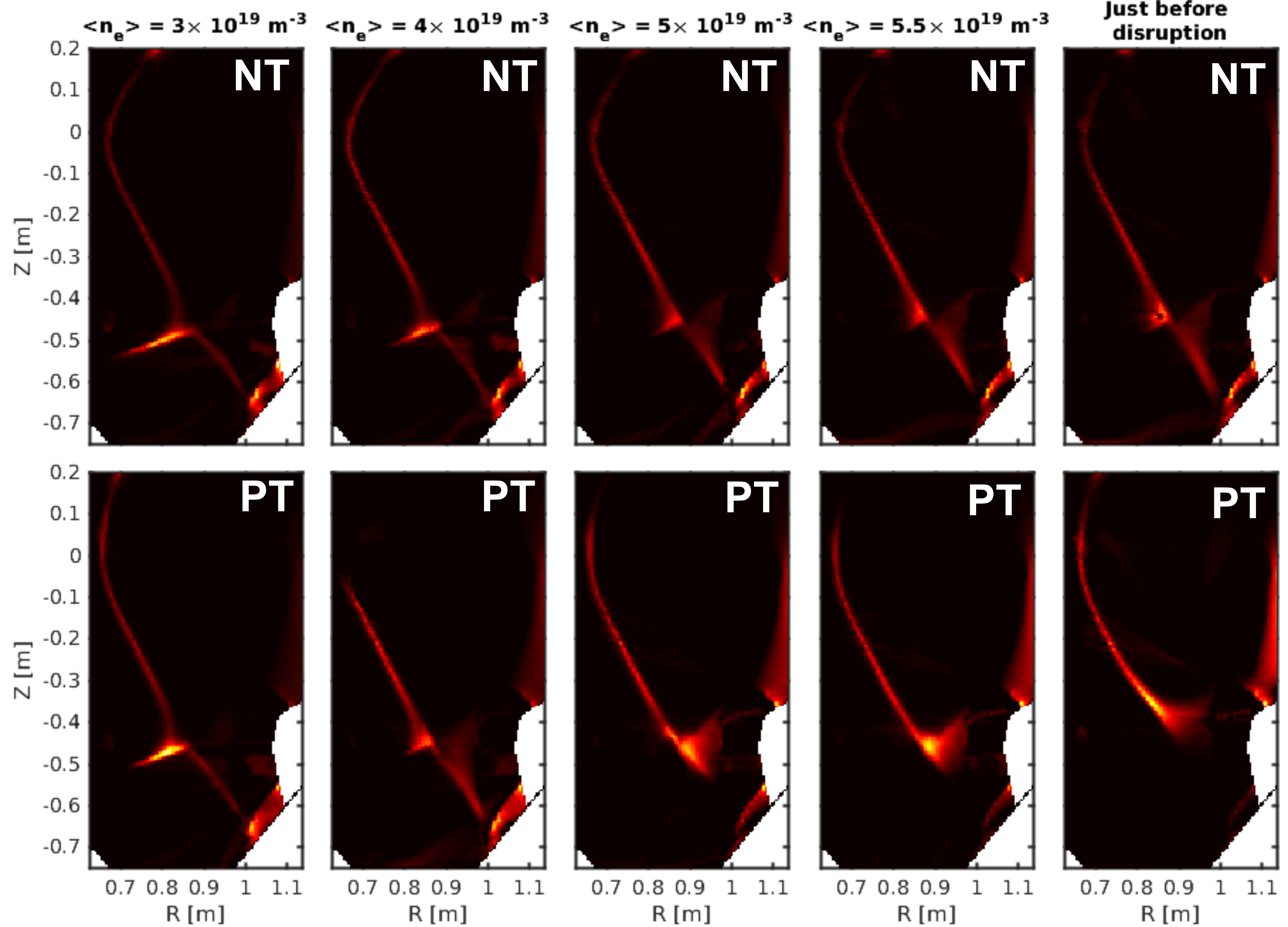}
\caption{\label{fig:CIII_Inversions_66957_vs_66962} Tomographic inversions of the filtered (CIII) images of the divertor at different \NE\ for the cases investigated in section \ref{sec:role_divertorshape} in NT (top line) and PT (bottom line). The colormap is the same for all panels.}
\end{figure*}

We now employ a different indicator of detachment, the CIII emission location. It was shown in previous TCV studies that the position of the CIII emission along a divertor leg provides a convenient tool to assess the detached status of the divertor \cite{Harrison_NME2017,Theiler_NF2017, Fevrier_NME2021}. Because of its strong dependence upon the local electron temperature, the CIII emission location was found to be a good indicator of the position of a low temperature region along the divertor leg. Figure \ref{fig:CIII_Inversions_66957_vs_66962} plots tomographic inversions of CIII divertor images taken from the MANTIS diagnostics \cite{Perek_RSI2019}. Unfortunately, the inversions exhibit an artifact in the vicinity of the outer strike point. Despite this, the CIII radiation front remains close to the NT target, while it separates from the target for PT. Concentrating upon the inner leg, both cases show a movement of the CIII front towards the X-Point, at comparable paces, indicating detachment for both triangularities. To conclude this section, it appears that changing the upper triangularity from positive to negative while keeping the lower triangularity and divertor shapes identical still exhibits a harder access to outer target cooling.

\subsection{Influence of triangularity on the SOL width}\label{sec:lambda_q}
Some experimental evidence suggested that NT configurations have a lower SOL width ($\lambda_q$) than PT configurations \cite{Faitsch_PPCF2018}. Recent numerical simulations also obtained this trend \cite{Muscente_NME2023,Lim_PPCF2023}. A lower $\lambda_q$ will result in a more concentrated heat flux, that, in turn, makes detachment more challenging. The SOL width is estimated using an Eich-fit function \cite{Wagner_NF1985,Eich_NF2013}, estimating $\lambda_q$, which quantifies the width of the SOL at the upstream location, and $S$, the spreading factor that quantifies the effect of perpendicular divertor transport on the broadening of the target profiles. It is then possible to determine the integral decay width, $\lambda_{int} \approx \lambda_q + 1.64 S$, that quantifies the radial extent over which power is deposited on the target plates \cite{Eich_JNM2013}. We now attempt to quantify how changes in SOL width may be responsible for the observed discrepancy between PT and NT detachment. 

\begin{figure*}[ht!]
\centering
\includegraphics[width=\linewidth]{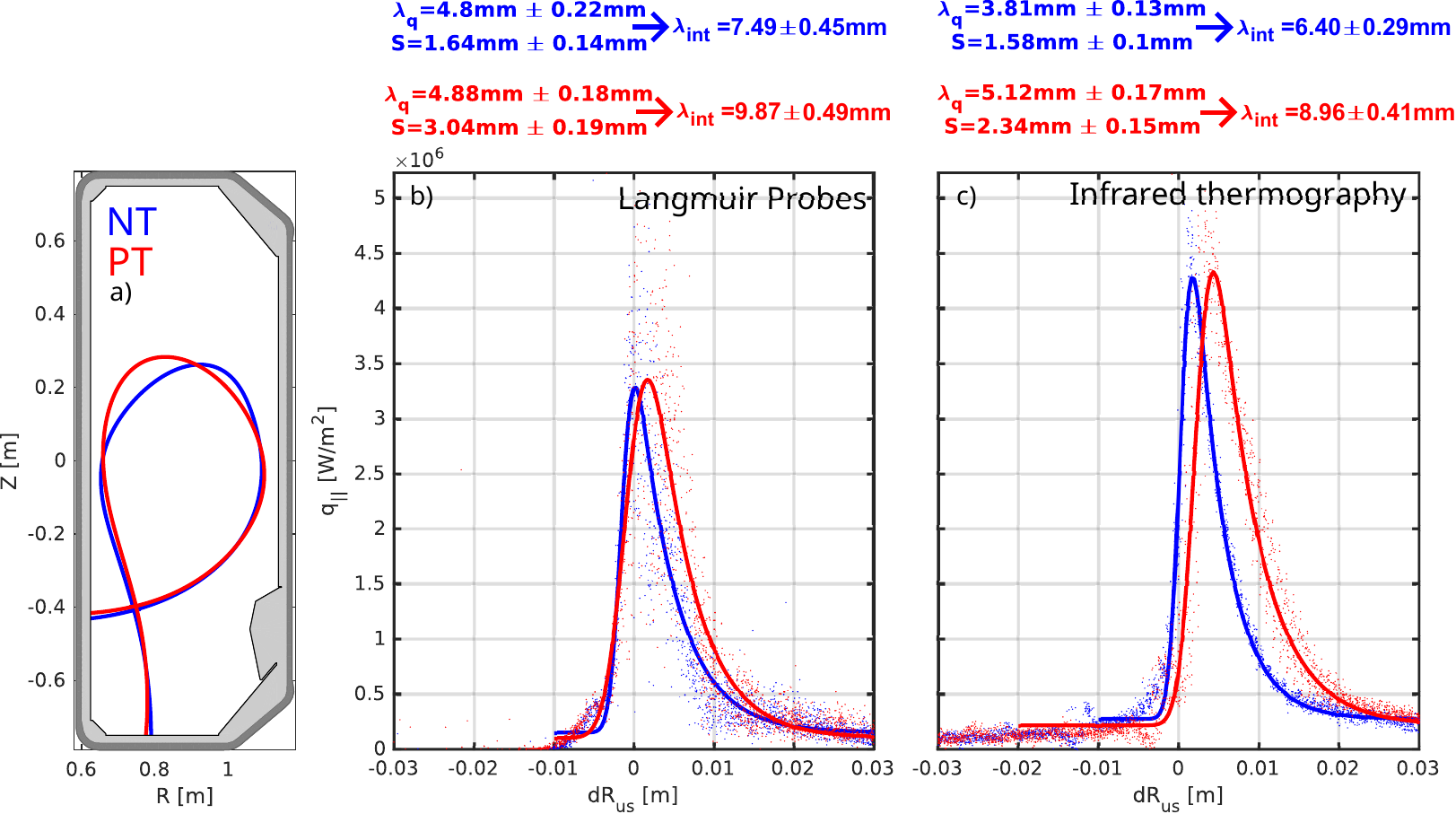}
\caption{\label{fig:lq_top_delta_scenario} (a) Shape of the NT (blue, \#74571) and PT (red, \#74413) equilibria considered in section \ref{sec:lambda_q}. (b) Parallel heat flux profiles, remapped to the upstream distance to the separatrix, dR$_{us}$, using LP and the procedure described in Ref. \cite{Gorno_PPCF2023} to account for non-ambipolar conditions. (c) Parallel heat flux profiles inferred from Infra-red thermography, remapped to the upstream distance to the separatrix, $dR_{us}$.}
\end{figure*}

We start by investigating the role of $\delta_{top}$, Figure \ref{fig:lq_top_delta_scenario}a. The outer strikepoint parallel heat-flux is determined from both Langmuir Probes (LP) using the methodology developed in Ref. \cite{Gorno_PPCF2023}, Figure \ref{fig:lq_top_delta_scenario}b, and Infra-Red thermography (IR), Figure \ref{fig:lq_top_delta_scenario}c. For both diagnostics, NT and PT show a similar peak parallel heat flux, whilst the absolute peak value differs by approximately ~20\% between the diagnostics. The data are fitted with an Eich-fit function \cite{Eich_NF2013,Wagner_NF1985}. While the LP-inferred $\lambda_q$ is similar for NT and PT, the IR inferred $\lambda_q$ is lower in the NT case. Conversely, the inferred spreading factor $S$ from both diagnostics is lower for NT ($46\%$ lower from LPs, $32\%$ lower from IR), leading to a lower $\lambda_{int}$. This is compatible with \cite{Faitsch_PPCF2018, Muscente_NME2023,Lim_PPCF2023}, although these results tend to attribute the lower SOL width to a smaller $S$ in NT, rather than a smaller $\lambda_q$. Due to the uncertainties and assumptions entering the analysis of both diagnostics, as well as the uncertainties associated with the fits, these results do not exclude a $\lambda_q$ smaller or equal in NT than PT.

\begin{figure}[ht!]
\centering
\includegraphics[width=\linewidth]{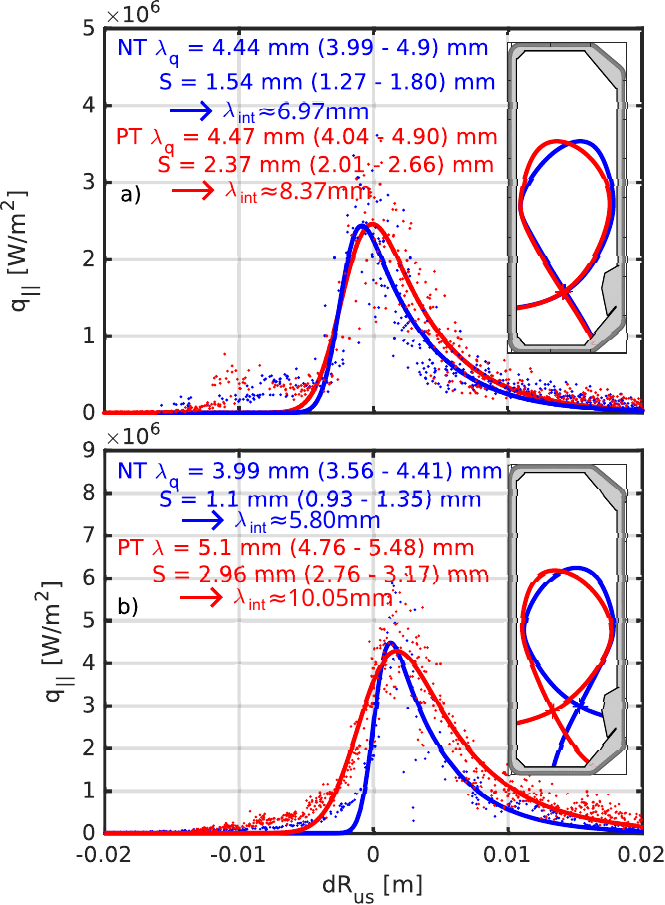}
\caption{\label{fig:lq_hybrid_scenario} Parallel heat flux profiles, remapped to the upstream distance to the separatrix, dR$_{us}$, using LP and the procedure described in Ref. \cite{Gorno_PPCF2023} to account for non-ambipolar conditions. The numbers in parenthesis indicate the 95\% confidence interval from the fit, with the NT (blue) and PT (red) shapes considered plotted in the inset. Two sets of cases are analyzed: (a) Discharges considered in section \ref{sec:role_divertorshape}. (b) Discharges taken from section \ref{sec:detachment}, Table \ref{tab:shotlist_densityscan}.}
\end{figure}

We now compare the inferred $\lambda_q$ for two of the discharges studied in section \ref{sec:role_divertorshape}, Table \ref{tab:hybrid_shapes_ff}. Figure \ref{fig:lq_hybrid_scenario}a (inset) plots the investigated geometries together with the parallel heat flux profiles, determined from LP, at the outer strike-point. $\lambda_q$ is measured in the attached phase of these discharges, before the \NE\ ramp, with the NT discharge at $\left<n_e\right>\approx 3\times 10^{19}~\mathrm{m}^{-3}$ and the PT $\left<n_e\right>\approx 2.5\times 10^{19}~\mathrm{m}^{-3}$. Consistently with the observations of Figure \ref{fig:lq_top_delta_scenario}b with LPs, both NT and PT have the same $\lambda_q$, while $S$ is $\approx33\%$ lower for NT, again corresponding to a lower $\lambda_{int}$ for NT. We now investigate the role of the bottom triangularity, for two of the discharges studied in section \ref{sec:role_divertorshape}, Table \ref{tab:hybrid_shapes}. Figure \ref{fig:lq_hybrid_scenario}b plots the investigated geometries together with the parallel heat flux profiles, determined from LP, from the outer strike-point. A 20\% lower $\lambda_q$ is obtained for the NT as compared to PT. Again, we find a lower $S$ in the NT plasma ($\approx63\%$), leading to a lower $\lambda_{int}$. 

Taken together, these experimental measurements indicate that negative triangularity is associated with narrower strike-points, in agreement with previous studies. However, our results suggest a role of the triangularity on $S$ rather than $\lambda_q$, disagreeing with previously published works and modelling evidence. This calls for further investigations and multi-machine comparison. Narrower heat flux profiles, for a given target geometry and total heat flux, would lead to higher peak heat flux, making detachment harder to attain. However, in cases where target geometry and connection lengths are similar, as in Figure \ref{fig:lq_hybrid_scenario}a), detachment remains difficult for NT configurations, even for comparable heat flux profiles. In the next section, we show how this may be related to a change in the overall particle balance of these discharges when changing the upper triangularity. 


\subsection{The role of fuelling and divertor neutral pressure}\label{sec:divertor neutral pressure}
\subsubsection{Investigating the role of the fuelling strategy}\label{sec:subsub_fuellingstrategy}
\begin{figure}[ht!]
\centering
\includegraphics[width=\linewidth]{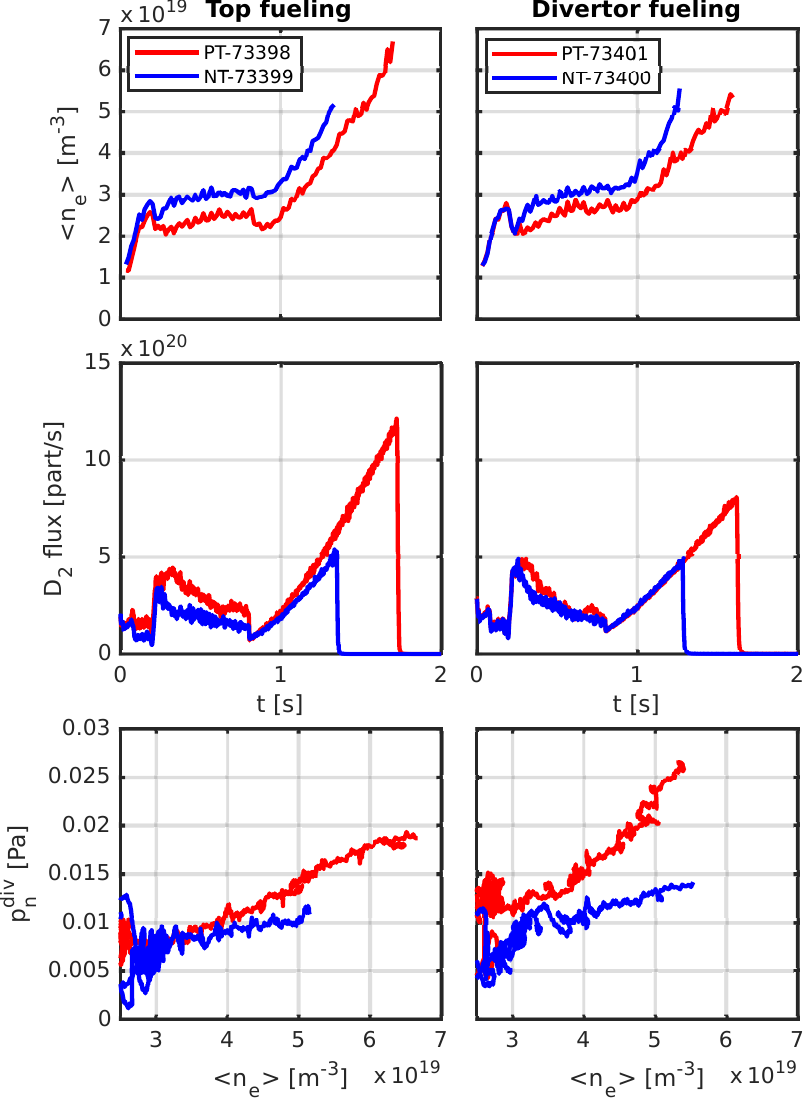}
\caption{\label{fig:neutral_pressure_hybrid} (top) Time evolution of \NE\ and (middle) the $D_2$ gas flux for the NT and PT cases of Table \ref{tab:hybrid_shapes_ff} and Table \ref{tab:hybrid_shapes}, left for top fuelling and right for divertor fuelling. (bottom) Divertor neutral pressure $p_n^{div}$ versus \NE\ for the NT and PT cases of Table \ref{tab:hybrid_shapes_ff} and Table \ref{tab:hybrid_shapes}, left for top fuelling and right for divertor fuelling.}
\end{figure}

To investigate the impact of the fuelling strategy (feedback controlled), we conducted a series of experiments using programmed fuelling. We additionally investigated differences between divertor and main chamber fuelling. The geometries used are similar to those shown in Figure \ref{fig:Hybrid_summary}. The scenarios are fuelling ramps with a plasma current $I_P = 225~\mathrm{kA}$ and the magnetic field in the favorable $\nabla B$ direction. Table \ref{tab:hybrid_shapes_ff} shows a summary of the discharges, and Figure \ref{fig:neutral_pressure_hybrid} plots the corresponding $D_2$ fuelling rates, together with \NE\ from TS measurements and the divertor neutral pressure $p_n^{div}$ measured by a floor positioned baratron. For a given $D_2$ influx, \NE\ is higher for the NT configurations and $p_n^{div}$ is systematically lower for NT case. Such a difference in neutral pressure could have been expected following observations of a lower outer target particle fluxes (Figure \ref{fig:LP_OSP_integrated}) for NT, as the main source of neutrals in the divertor comes from the recycling of the ions impinging upon the divertor targets. Yet, this, together with observation of a higher core density for the same fuelling rate, indicates a different particle balance for NT vs PT configurations, likely linked to a higher particle confinement for NT. 
\begin{table}[h]
\center
\begin{tabular}{ l c c c c }
\hline
Discharge& $\delta_{top}$ & $\delta_{bot}$ & fuelling location\\
\hline
\textcolor{red}{73398} & 0.19 & 0.05 & V2 (Top) \\
\textcolor{blue}{73399} & -0.29 & 0.03 & V2 (Top) \\
\textcolor{blue}{73400} & -0.29 & 0.04 & V1 (Divertor) \\
\textcolor{red}{73401} & 0.2 & 0.05 & V1 (Divertor) \\
\hline
\end{tabular}
\caption{{\label{tab:hybrid_shapes_ff} Summary of the main plasma discharges used in section \ref{sec:subsub_fuellingstrategy}. The geometries used are similar to the ones shown in Figure \ref{fig:Hybrid_summary}.}}
\end{table}

\subsubsection{Observation of different divertor neutral pressure in NT vs PT}\label{sec:subsub_divertorneutralpressure}
\begin{figure}[ht!]
\centering
\includegraphics[width=\linewidth]{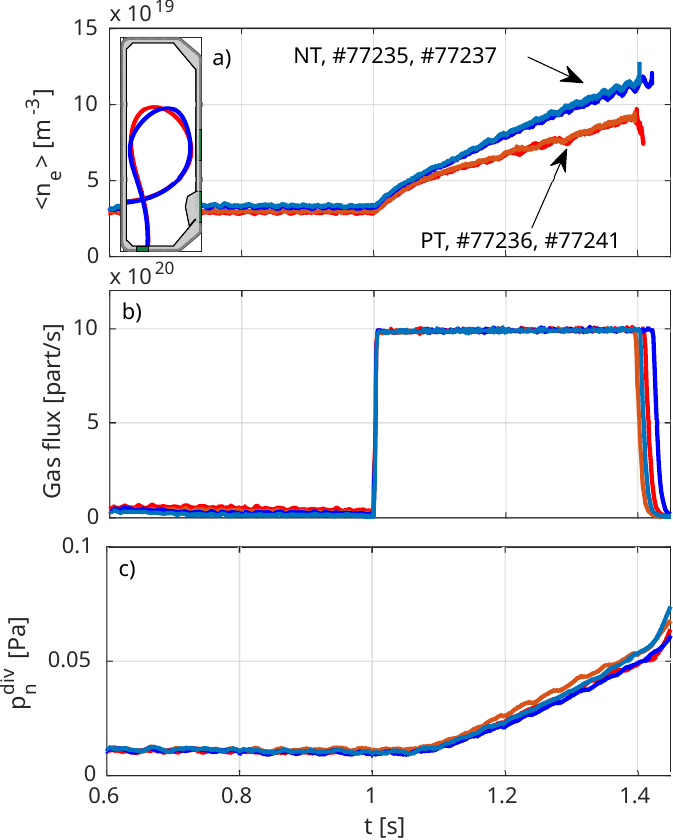}
\caption{\label{fig:divertor_pressure_scan_timeevoution} Evolution of (a) the line-averaged density \NE, (b) the D$_2$ gas flux, (c) the divertor neutral pressure for PT and NT cases listed in table \ref{tab:divertor_pressure_scan}. The inset of panel (a) plots the (blue) NT equilibrium (top-$\delta\approx -0.23$, bot-$\delta\approx 0.59$) (red) PT equilibrium (top-$\delta \approx 0.25$, bot-$\delta\approx 0.59$).}
\end{figure}

\begin{figure}[ht!]
\centering
\includegraphics[width=\linewidth]{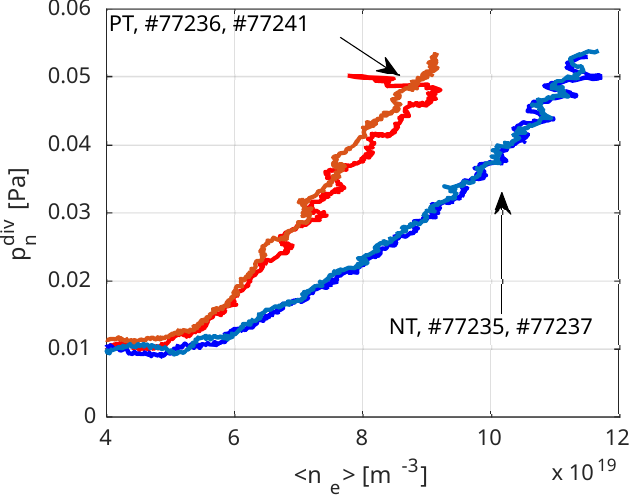}
\caption{\label{fig:divertor_pressure_scan_pndiv} Evolution of the divertor neutral pressure as a function of \NE, for PT and NT cases listed in table \ref{tab:divertor_pressure_scan}, section \ref{sec:divertor neutral pressure}.}
\end{figure}

In section \ref{sec:subsub_fuellingstrategy}, discharges with matched divertor shape revealed that the divertor neutral pressure, $p_n^{div}$, measured from a floor baratron, Figure \ref{fig:neutral_pressure_hybrid}, was lower in NT than PT. Therefore, we conducted an experiment to examine the role of $\delta_{top}$ on $p_n^{div}$. These discharges featured a positive $\delta_{bot}\approx 0.59$ similarly to the discharges examined in \cite{Faitsch_PPCF2018}, Figure \ref{fig:divertor_pressure_scan_timeevoution}a (inset). Particular attention was paid to the plasma geometry, such as the proximity to the (carbon) wall, which can impact the particle recycling. To reduce effects of machine history, the discharges were performed sequentially and employed fixed fuelling rates to prevent any complexities engendered by feedback control. Table \ref{tab:divertor_pressure_scan} provides an overview of the discharges discussed in this section, and Figure \ref{fig:divertor_pressure_scan_timeevoution} plots the time evolution of \NE, the D$_2$ gas flux, and $p_n^{div}$. The fuelling rate was chosen at $1\times 10^{20}$ part/s, strong enough to achieve a density limit disruption. 
\begin{table}[ht]
\center
\begin{tabular}{ l c c c c }
\hline
Discharge& Configuration & $\delta_{top}$ & fuelling rate \\
\hline
\textcolor{blue}{77235} & NT & -0.23 & $1\times 10^{20}$ part/s \\
\textcolor{red}{77236} & PT & 0.25 & $1\times 10^{20}$ part/s \\
\textcolor{matlabblue}{77237} & NT & -0.23 & $1\times 10^{20}$ part/s \\
\textcolor{matlabred}{77241} & PT & 0.25& $1\times 10^{20}$ part/s \\
\hline
\end{tabular}
\caption{{\label{tab:divertor_pressure_scan} Summary of the main plasma discharges used in section \ref{sec:divertor neutral pressure}.}}
\end{table}
Figure \ref{fig:divertor_pressure_scan_pndiv} plots the evolution of the divertor neutral pressure as a function of \NE\ for the discharges in table \ref{tab:divertor_pressure_scan}. For identical \NE, the NT discharges exhibit a lower $p_n^{div}$, in agreement with the observations from section \ref{sec:subsub_fuellingstrategy}. This result also holds after accounting for the finite time delay in neutral pressure measurements by the baratron gauges ($\approx 100~\mathrm{ms}$). Interestingly, the NT discharges reach a higher \NE\ before disrupting, which is consistent with a MARFE-triggered disruption. 
As $p_n^{div}$ is an essential parameter governing the access to divertor detachment, these observations explain, at least in part, the reduced detachment access for NT. Together, all these observations indicate a change in the particle balance between NT and PT configurations. However, physical processes responsible for this difference remain to be explained and are left for future work. 

\section{Conclusion} \label{sec:conclusion}
In this article, we explored experimentally detachment physics in negative triangularity configurations. From the power exhaust viewpoint, negative triangularity has several advantages. Operating only in L-mode removes any need for ELM mitigation or ELM buffering. The power crossing the separatrix is not required to remain above any L-H threshold, allowing for higher core and edge radiation. It remains necessary to characterize detachment in such configurations. These TCV experiments revealed a surprising difficulty to attain detachment in Ohmic NT configurations, compared to their PT equivalents, with no target plasma cooling below the typical 5 eV detachment threshold achieved in core density ramps, for favorable and unfavorable field directions. Several candidate effects to explain these observations were investigated. No significant difference between PT and NT in terms of the upstream density and power going to the SOL was found. Geometrical divertor effects, such as the role of the poloidal leg length, of the parallel connection length, and of the ratio of connection length between inner and outer targets, were studied. While some of these effects were important, they could not, alone, account for the difference in detachment behavior, as shown by experiments with matching divertor shapes. We investigated the role of $\lambda_q$ in these observations, known to reduce detachment access for decreasing $\lambda_q$. Previous studies indicated that $\lambda_q$ was smaller for NT configurations \cite{Faitsch_PPCF2018, Muscente_NME2023, Lim_PPCF2023}. LP measurements indicated that target heat flux profiles are narrower, in terms of $\lambda_{int}=\lambda_q+1.64S$, for NT than PT. These measurements attributed the difference to a reduction of the spreading-factor $S$, rather than of $\lambda_q$, in disagreement with previously published works and modelling evidence. This calls for further investigation, in particular in explaining the experimental discrepancy between heat-fluxes inferred from IR and LP. Finally, for discharges with matched divertor geometries and variation of the upper triangularity, we noted that, for a given $\mathrm{D}_2$ fuelling flux, the line-averaged density was typically higher in NT configurations. Furthermore, for matched line-averaged densities, the divertor neutral pressure was typically lower in the NT configurations, which explains in part why they are harder to detach. This suggests that not only energy confinement, but also particle confinement, are affected by triangularity.

From the core perspective, NT remains an exciting no-ELM regime for a future DEMO reactor. However, a more complete exhaust characterization is still required, and our investigations suggest that multiple factors may impact detachment for NT. Further studies are needed to explore these differences. Continual improvements in TCV's diagnostic coverage will help shed light on this issue. While this work focused on detachment with core density ramps, detachment with impurity seeding will be explored in the future. Preliminary results, with Nitrogen, indicate that detachment can then be achieved for NT, but at the expense of reduced core confinement. We also intend to explore NT detachment with high-input power using TCV's ECH and NBH systems. Indeed, to validate the NT tokamak concept from the exhaust point of view, one needs to demonstrate the operation of a fully detached L-mode negative triangularity plasma with performances that match those of an H-mode PT plasma, likely to be only inter-ELM detached.

\section*{Acknowledgements}
This work has been carried out within the framework of the EUROfusion Consortium, partially funded by the European Union via the Euratom Research and Training Programme (Grant Agreement No 101052200 — EUROfusion). The Swiss contribution to this work has been funded by the Swiss State Secretariat for Education, Research and Innovation (SERI). Views and opinions expressed are however those of the author(s) only and do not necessarily reflect those of the European Union, the European Commission or SERI. Neither the European Union nor the European Commission nor SERI can be held responsible for them. This work was supported in part by the Swiss National Science Foundation. This work was supported in part by the US Department of Energy under Award Number DE-SC0010529.

\section*{References}
\bibliographystyle{unsrt}
\bibliography{OFevrier_NegDelta}
\end{document}